\documentclass{article}
\usepackage{graphicx}
\usepackage[utf8]{inputenc}
\usepackage{mathtools}
\usepackage{lineno,hyperref}
\usepackage{amsmath,mathtools}
\usepackage{multicol}
\usepackage{multirow}
\usepackage{amssymb}
\usepackage{subfig}
\usepackage{pdflscape}
\usepackage{listings}
\usepackage{authblk}
\usepackage{algorithmicx,algpseudocode}
\usepackage[ruled,vlined,linesnumbered]{algorithm2e}
\usepackage{biblatex} 

\providecommand{\keywords}[1]{\textbf{\textit{Keywords---}} #1}

\begin{document}

\title{A Composite Centrality Measure for Improved Identification of Influential Users~\footnote{Preprint submitted for publication}}
 
\author{ Ahmad Zareie$^1$ (ahmad.zareie@postgrad.manchester.ac.uk) \and Amir Sheikhahmadi$^2$ (sheikhahmadi@eng.ui.ac.ir) \and
Rizos Sakellariou$^1$ (rizos@manchester.ac.uk)}

\affil{$^1$Department of Computer Science, The University of Manchester, Manchester M13 9PL, U.K.}
\affil{$^2$Department of Computer Engineering, Sanandaj Branch, Islamic Azad University, Sanandaj, Iran}
\date{}
\maketitle

\begin{abstract}
In recent years, the problem of identifying the spreading ability
and ranking social network users according to their influence has attracted a lot of
attention; different approaches have been
proposed for this purpose. Most of these approaches rely on the topological location of nodes and their neighbours in the graph to provide a measure that estimates the spreading ability of users. One of the most well-known measures is k-shell; additional measures have been proposed based on it. However, as the same k-shell index may be assigned to nodes with different degrees, this measure suffers from low accuracy.
This paper is trying to improve this by proposing a composite centrality measure in that it combines both the degree and k-shell index of nodes. Experimental results and evaluations of the proposed measure on various real and artificial networks show that the proposed measure outperforms other state-of-the-art measures
regarding monotonicity and accuracy.
\end{abstract}


%\end{keyword}

%\tableofcontents

% ************************************************************************************

\section{Introduction}
\label{Section:1}
Social networks have recently received lots of attention as an important means to capture the interactions of people’s activities, including their communications, preferences and so on.
%marketing, entertainment and so on, 
%and the number of their users is on increase day to day. 
The number of users and the diversity of data that can be exchanged in such networks have motivated research into fast and efficient ways to spread news, messages as well as marketing advertisements~\cite{tarikh2}. 
In this model, which is commonly referred to as viral marketing~\cite{tarikh1,G3}, the aim is to carefully select a small number of users and help spread messages in the network through them~\cite{G4,FrogIM}.
The key idea is to avoid the cost and overhead of sending messages to all users. The popularity of the viral marketing model relies on properties such as the trust between social friends and easier acceptance of messages by friends~\cite{T4}.
%given the
%impossibility of connection to all users and sending messages
%to them due to expensive costs and time overhead, a small
%number of users are chosen and messages are spread in the
%network with the help of them  The trust  also is
%another advantage of viral marketing which increases the
%popularity of the model among companies and firms .

For one, all users are not equally influential. Some users can spread
messages more widely due to their topological location and/or number of 
friends (e.g., connected nodes) in the network. The identification of the users' spreading
ability and the selection of a set of influential users for message
spreading is one of the most important challenges in the
model. Proposing an appropriate and accurate measure to capture the perceived influence of users, so that they can be ranked according to how influential they are perceived to be, is particularly important for viral marketing success~\cite{T3}. In existing research, most of these measures are based on the notion of centrality~\cite{Survey}, which takes into account the topological location of users and their friends.

One of the most widely used centrality measures is k-shell~\cite{kshell}, which is based on the observation that the closer to the core~\cite{core} of a network a node is, the more influential it should be. 
In k-shell decomposition, an algorithm eliminates the nodes of the network in some orderly manner (steps), and a k-shell index is assigned to each node based on the step in which the node is eliminated. The nodes with greater indices are closer to the core of the network and
are accordingly more influential. Given the suitable accuracy
and efficiency of the k-shell method in comparison to other centrality measures, such as closeness~\cite{closeness}, betweenness~\cite{betweenness}, and degree~\cite{degree}, lots of work has focused on improving the calculation of the k-shell index~\cite{KSsurvey}. Some of this work includes coreness~\cite{Cnc+}, k-shell iteration factor~\cite{kshell-if}, mixed degree decomposition~\cite{MDD}, hierarchical k-shell~\cite{HKS}, k-shell hybrid~\cite{Improved_KShell_Hybrid}, k-shell based on gravity centrality~\cite{KSGravity}, improved k-shell~\cite{ImprovedKS} and neighbourhood diversity~\cite{EDSR}. 
In k-shell decomposition, the same k-shell index is assigned to nodes with a different degree value, something that decreases the accuracy of the method. This is because nodes with the same k-shell but different degree do not have the same spreading ability; 
nodes with a larger degree can spread messages more widely compared to nodes with a lower degree in the same k-shell.
Although iterations in each step have been considered to improve the accuracy of k-shell by some of the mentioned measures~\cite{kshell-if,HKS}, there is scope for further improvement.

In this paper, we propose an improved composite centrality measure to identify and rank influential nodes, which builds upon the strength of k-shell and makes use of covariance to combine k-shell with degree. Thus, the spreading ability of a node is calculated according to the centrality of its neighbours. This allows us to develop a more precise centrality measure to differentiate nodes, which outperforms other centrality measures as demonstrated by a set of comprehensive experiments.
%differentiates nodes spreading abilities and its accuracy in ranking the nodes are investigated using comprehensive experiments, and the obtained results are compared to the other methods.

The rest of the paper is organized as follows. Section~\ref{Section:2} reviews relevant previous studies. The motivation behind the proposed method is described in Section~\ref{Section:3}. The proposed method is explained in detail in Section~\ref{Section:4}. Experimental results are reported in Section~\ref{Section:5}. Finally, Section~\ref{Section:6} summarizes the paper and gives some directions for future work.

%***********************************************************************************************************************************************************************
\section{Related Work}
\label{Section:2}
Identifying and ranking users' spreading ability
has attracted lots of attention in recent years and different
centrality measures have been proposed. In these measures, the spreading ability of nodes is determined based on their
topological features and network structure. Degree centrality~\cite{degree}, which identifies the nodes' influence based on the number of their neighbours, is one of the simplest methods. 
In~\cite{DIL}, the importance of each link is determined based on the number of  triangles containing the link; degree and importance of the links connected to each node are used to propose a centrality measure.
In~\cite{H-index}, a measure is proposed to identify nodes' influence and rank them, based on the degree of the neighbours and the h-index concept. Various methods have also extended the h-index concept in~\cite{EHC,localH-index}. Entropy centrality~\cite{ERM} is another method which determines nodes' spreading ability based on their first- and second-order neighbours' degree. In clustered local-degree~\cite{CLD}, nodes' influence is estimated based on the degree of the neighbours and the clustering coefficient.

Given the closeness to the core of a network graph, k-shell
decomposition~\cite{kshell} divides the nodes into several shells and estimates their spreading ability based on these shells. In this method, in order to determine the closeness of nodes to the core, 1-degree nodes are removed from the graph leaving no node with degree 1. The removed nodes are considered in the
first shell, and a value of k-shell equal to 1 is assigned to them. In the next step, 2-degree nodes
are removed. The elimination of the nodes with degree 1 or 2 is repeated for as long as there is a node with degree equal to or less than 2. The removed nodes in this step are considered in the second shell, and a value of k-shell equal to 2 is assigned to them. By repetition of these steps, all nodes of the graph are removed and their k-shell is determined.  

Nodes with a higher k-shell value are located closer to the core and considered more influential. The summation of neighbours' k-shell is considered as a measure in~\cite{Cnc+} to determine nodes’ centrality. Nodes' degree and the iteration of the k-shell decomposition algorithm, in which nodes are removed from the graph, are taken into
account in~\cite{kshell-if} to estimate the spreading ability of nodes. According to the number of remained and removed neighbours
of nodes in different steps of the k-shell decomposition algorithm, an extension of k-shell centrality is proposed in~\cite{MDD}. The authors in~\cite{HKS} believe that apart from closeness to the core of graph, nodes’ distance to graph periphery is an important element to determine nodes' spreading ability. Thus, they propose a hierarchical extension of k-shell decomposition algorithm to determine influence spreading of nodes. In~\cite{CN} the neighbours of each nodes are first divided in four categories based on the iterations of k-shell decomposition in which nodes are removed. Then, nodes’ spreading ability is identified based on their neighbours in the four categories.

There has been work where compound centrality measures have been proposed. Degree, k-shell, and dispersion of the neighbours of nodes are taken into account to propose a compound centrality measure in~\cite{MCDE}; however, this is a measure that simply adds the different values and may not be accurate to distinguish the nodes with different degrees in a same shell.
%%----------------Changes-Ahmad
The same appears to be the case with the method proposed in~\cite{Improved_KShell_Hybrid}; this method combines  degree and k-shell of the nodes in the neighbourhood of a node to calculate the centrality of the node.  
%%---------------
Using the k-shell decomposition algorithm and entropy notations, spreading sphere and intensity of the neighbours of nodes are calculated in~\cite{EDSR}. Then, the nodes' spreading ability is determined based on sphere diversity and intensity of their neighbours. 
The number of neighbours in each hierarchy of shells is taken into account in~\cite{ECRM} to define the commonality between a node and its neighbours and improve the accuracy of clustering coefficient approach. A combination of degree and k-shell indices is proposed in~\cite{KSD} to determine the significance of each edge, and summation of the significance of the edges between each node and its neighbours is calculated as node spreading ability.
Edges significance and k-shell centrality are considered to propound an influence detection method in~\cite{LS}.
%-----------------Changes-Ahmad
In~\cite{ImprovedKS}, node information is calculated for each node using entropy notation; nodes in the same shell are sorted based on node information. In~\cite{KSGravity}, a method is proposed to calculate the attraction coefficient between each pair of nodes based on their k-shell; then the centrality of a node is determined based on the attraction coefficient, degree and distance of the node to the other nodes in the network. 
%-------------------

In contrast to previous work, our approach focuses on the neighbourhood of a node and the relationship between degree and k-shell to determine its centrality.
By considering the variation between shell and degree in the neighbourhood of a node, our approach  distinguishes more precisely than other measures the neighbours that are located in the same shell but have different degrees. This helps with the accurate determination of the neighbours' structural features.
Thus, the key novelty of our composite centrality measure is that it takes into account the joint variability of degree and k-shell, something that, as will be demonstrated in the experiments, leads to better discrimination and higher accuracy.
% ************************************************************************

\section{Motivation}
\label{Section:3}
\begin{figure*}[!tb]
 \center{
 
 \subfloat[EUR]{
   \centering
       \includegraphics[width=0.5\textwidth]{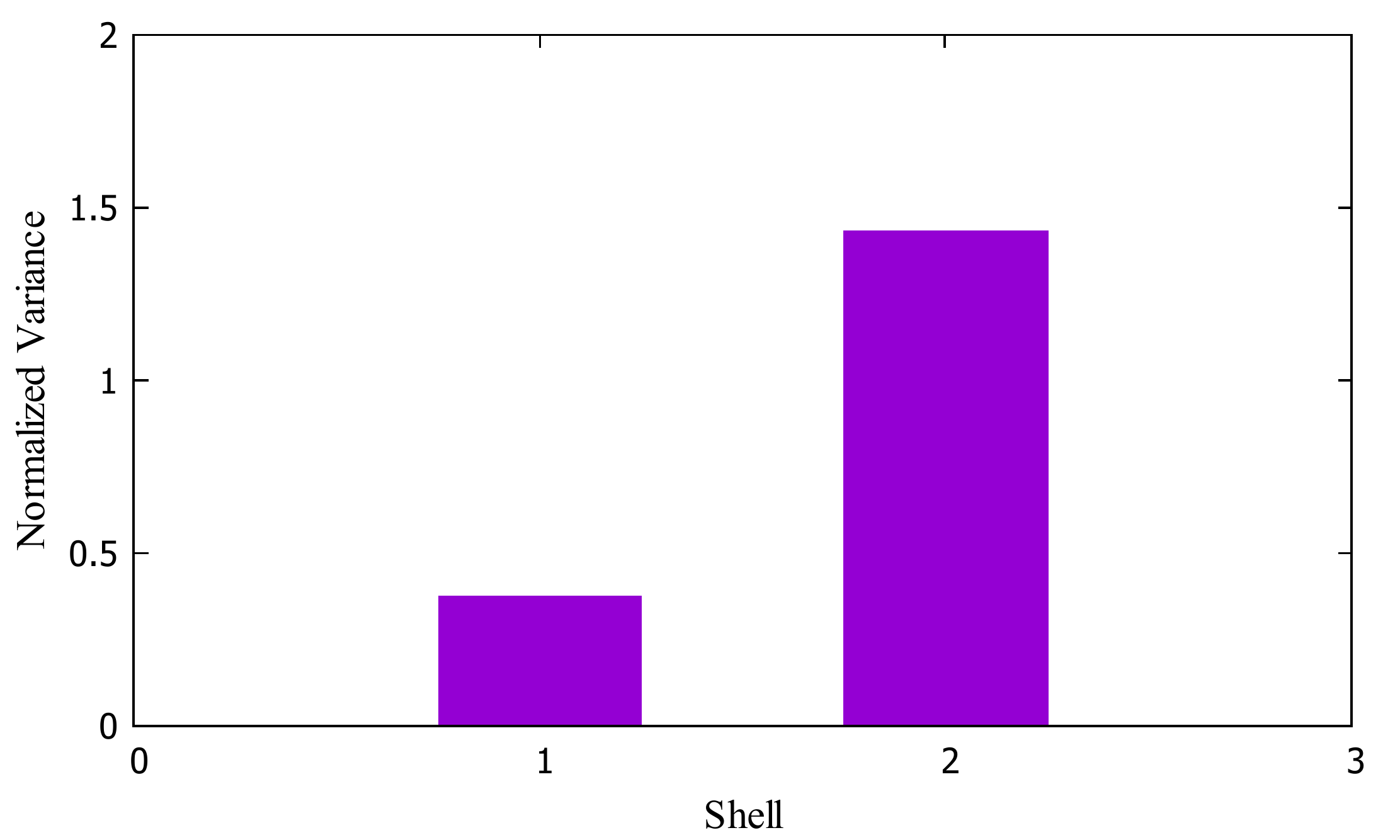}\label{Ana_EUR}
    }
 \subfloat[CHG]{
   \centering
       \includegraphics[width=0.5\textwidth]{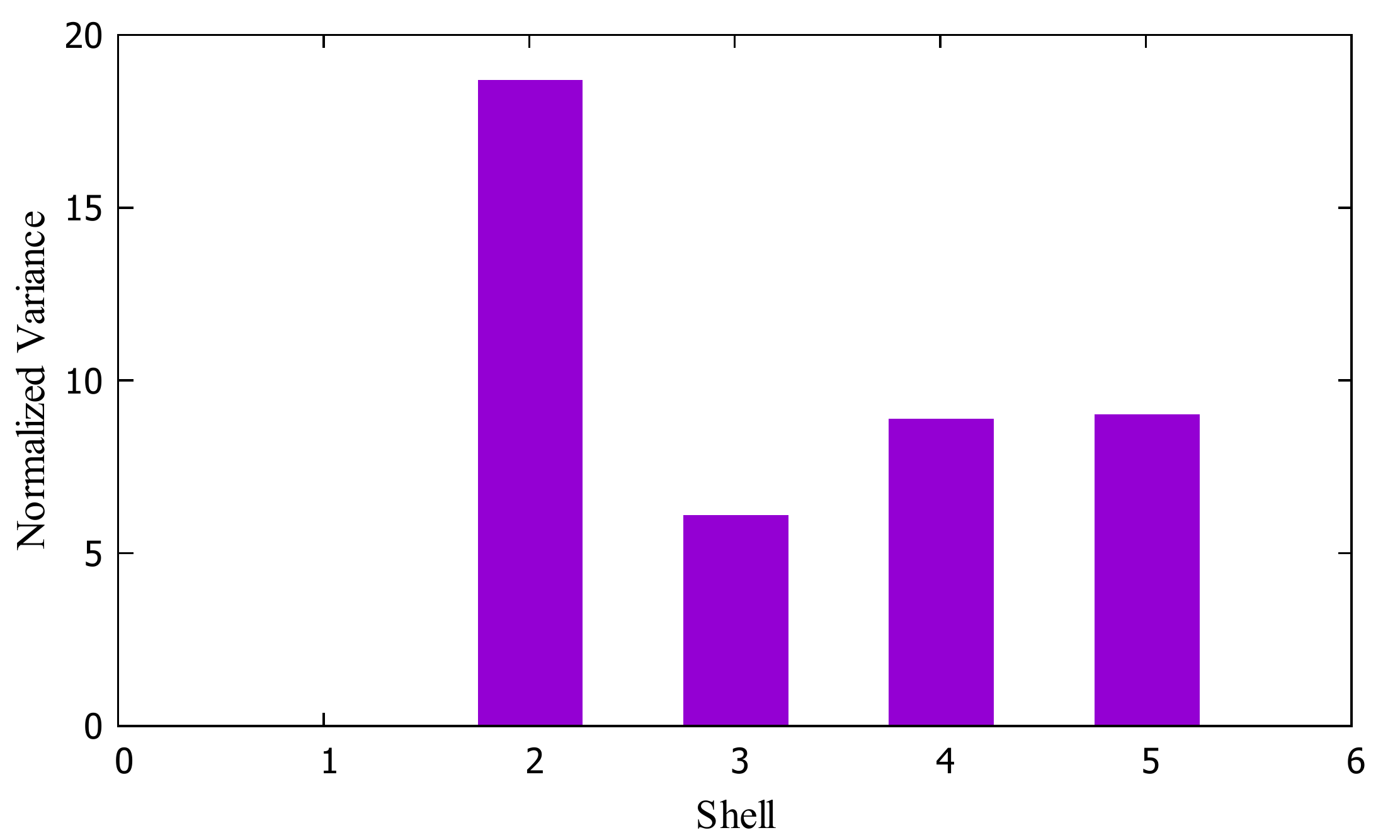}\label{Ana_CHG}
    }

 \subfloat[HMS]{
   \centering
       \includegraphics[width=0.5\textwidth]{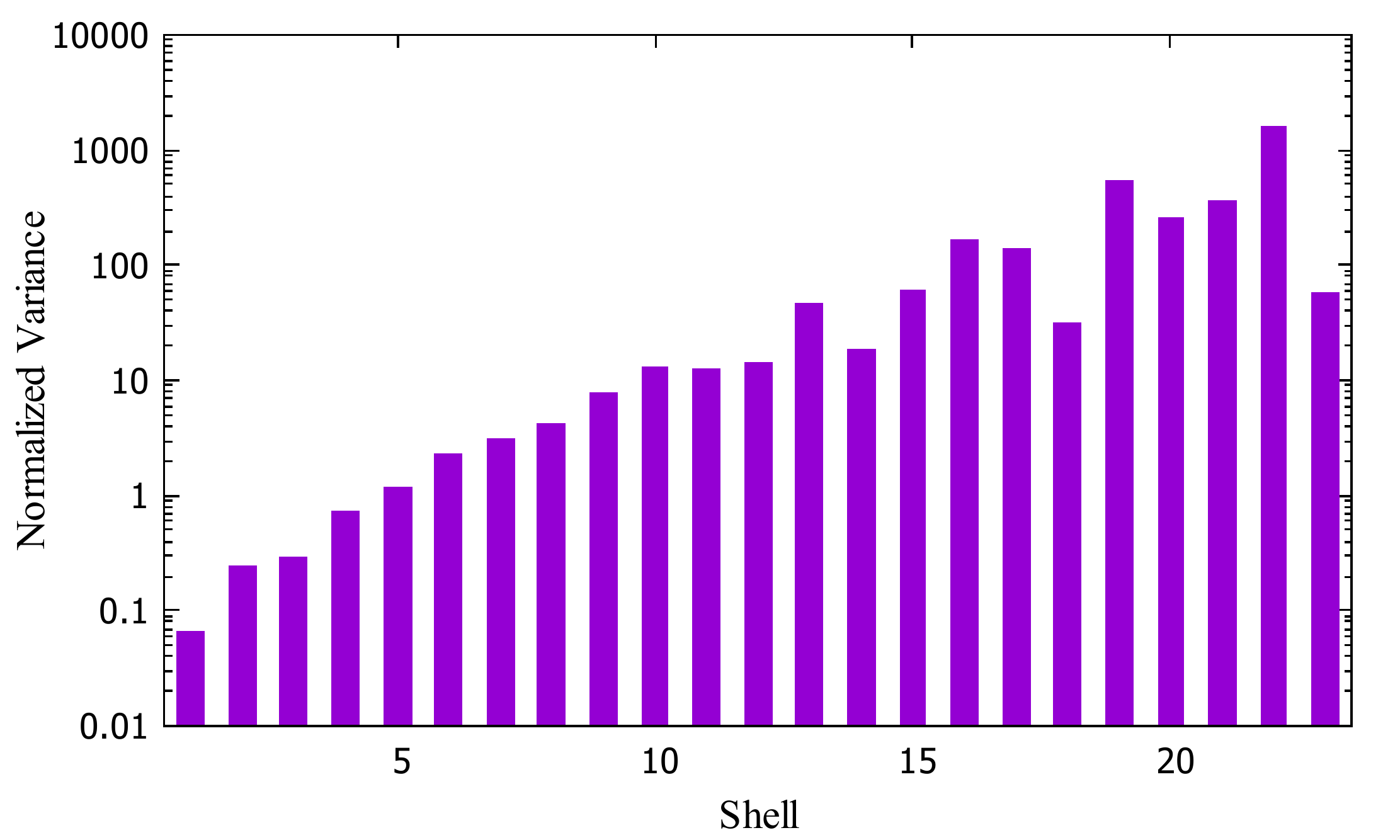}\label{Ana_HMS}
    }
        \subfloat[UPG]{
   \centering
       \includegraphics[width=0.5\textwidth]{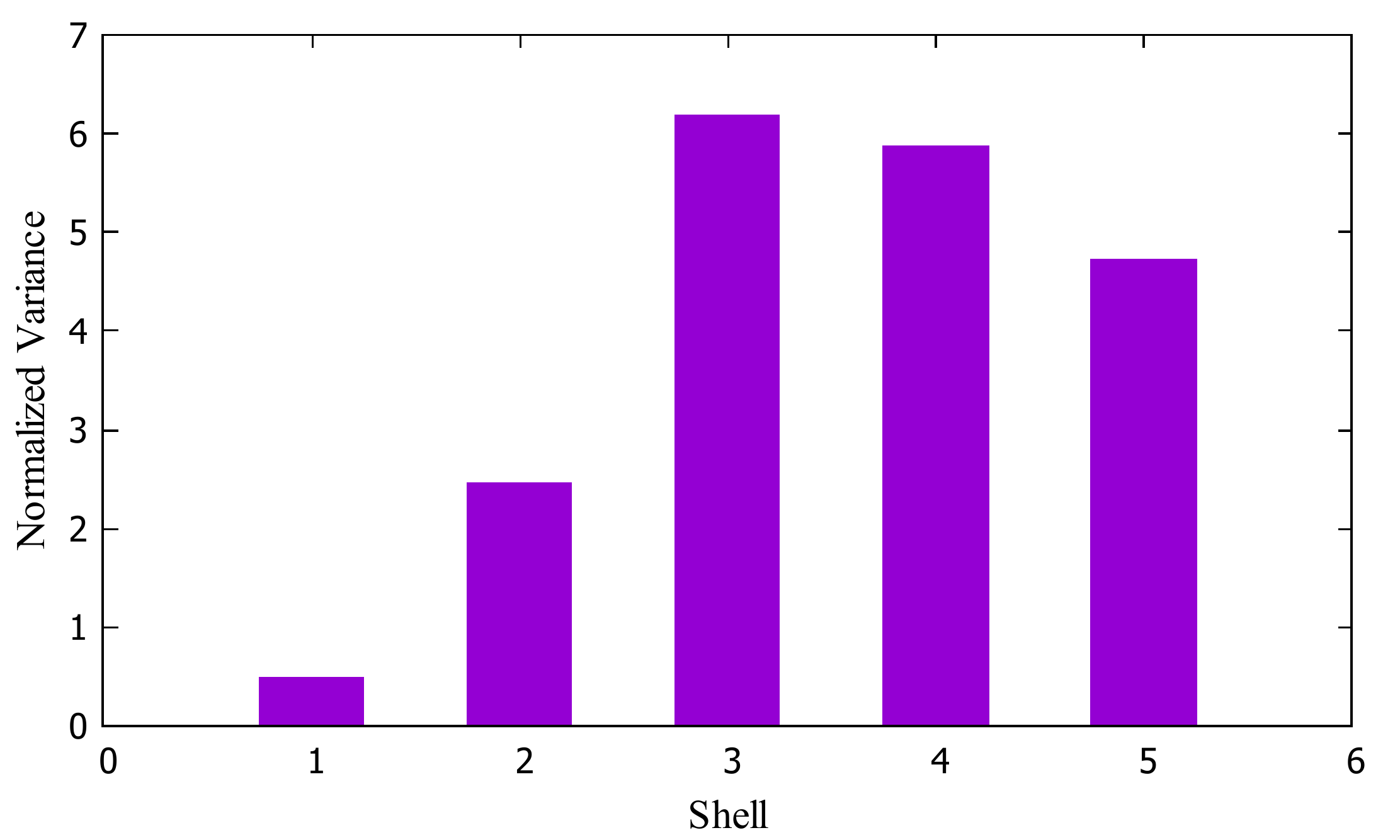}\label{Ana_UPG}
    }
       \caption{The normalized variance of the degree of the nodes in each shell for the EUR, CHG, HMS, UPG networks}
       \label{Fig_Analysis}
       }
 \end{figure*}

As mentioned in Section~\ref{Section:2}, k-shell centrality  is one of the best-known methods to determine nodes'
influence, and various methods have been proposed based
on it. Assigning the same k-shell index to the nodes with
different degrees can be regarded as one of the issues in
this centrality measure, as it leads to decreasing accuracy in nodes
ranking. To demonstrate this issue, four indicative real world networks, Hamsterster (HMS)~\cite{DatasetHAM}, Euroroad (EUR)~\cite{DatasetEUR}, US Power grid (UGP)~\cite{DatasetUPG} and Chicago (CHG)~\cite{DatasetCHG} have been considered. The purpose of this study was to investigate the degree diversity of the nodes in the same shell. The results of this study are shown in Fig.~\ref{Fig_Analysis}. The Normalized Variance of nodes' degree within each shell is reported in this figure.
The normalized variance is calculated as follows: if nodes $\{v_1,v_2,...,v_t\}$ are in a shell then the variance of the degree of nodes, i.e. $d_1,d_2,...,d_t$, is first calculated, then the result is divided by the number of nodes, $t$. 
%Then, the normalized variance in shell $f$ is calculated as   
%\[\frac{Var_{v1_1,v_2,\dots,v_t \in k-shell f}(d_1,d_2,\dots,d_t)}{t}\]
%This is calculated by Variance of the degree of the nodes in the shell per the number of the nodes in the shell.

% RIZOS - add a sentence to explain clearly how we see this from the figure
% Reply - normalized variance of the degree of nodes in each shell shows the variation of degree of the nodes in a same shell. For instance, in network HMS normalized variance in shell 22 is significant and shows the variation of the degree of nodes in this shell.
As seen in Fig.~\ref{Fig_Analysis}, the same k-shell is assigned to nodes with different degrees; the degree of the nodes in each shell varies significantly. For example, in the HMS network, the value of the normalized variance of the degree of nodes located in shell 22 shows that nodes with significantly different degree values are assigned in the same shell. This observation suggests that the k-shell decomposition method cannot determine accurately the structural information of nodes with different features.
% RIZOS - the following seems unnecessarily long - can you shorten?
% Reply - I removed 5 lines as it is now comment.
Although the nodes closer to the core of the graph are more influential in comparison to the peripheral nodes,
distinguishing the nodes located in each shell, based on their degree, can improve the ranking accuracy. 
%In other words, nodes whose neighbours are closer to the core of graph and have high degree can spread a message more widely. It is possible that the neighbours which are closer to the core have low degree and vice versa. In this situation considering degree or k-shell of the neighbours separately does not lead to an accurate estimation in the identification of nodes’ influentiality.
Thus, in order to determine the spreading ability of nodes, degree and k-shell of the neighbours are considered simultaneously in this paper. %and it is assessed if the neighbours being closer to the core have high degree. 
%That is to say, if degree and k-shell of the neighbours of a node are correlated, and if by increasing the k-shell in the neighbourhood of the node the degree increases.

In other words, the idea is that nodes having neighbours that are located in shells closer to the core of the graph and with large degree values are more central. By considering the variation of k-shell and degree in the neighbourhood of a node, this can help determine the structural features of neighbours and distinguish the nodes located in the same shell but having different degrees. In brief, the variation between degree and k-shell of the first and second-order neighbours of each node can be used as a basis of a covariance approach to centrality and motivates the proposed measure.

% *******************************************************************

\section{The Details of the Proposed Method}
\label{Section:4}
In this section the details of the proposed method are discussed. A social network can be modelled as a graph $G= (V, E)$. The sets $V=\{v_1,v_2,\dots,v_{|V|}\}$, representing users, and $E=\{e_1,e_2,\dots,e_{|E|}\}$, representing the relations between users, are the sets of nodes and edges of the graph, respectively. $N_z$ is the set of neighbours of node $v_z$. The cardinality of this set shows the degree of the node, $d_z$.

\subsection{Proposed Method}
In the proposed method, we first define a measure, called Shell-Degree-Centrality (SDC), to determine the variation between degree and k-shell of the first- and second-order neighbours of each node. The Covariance Centrality (CVC) of each node is then calculated based on the value of neighbours' SDC. Finally, the CVC of the neighbours is used to determine the Extended Covariance Centrality (ECVC) of each node, which indicates the spreading capability and rank of nodes in our proposed method. All this will be explained next.

In order to calculate the SDC of each node, two matrices $A^{(1)}(z)$ and $A^{(2)}(z)$ are first defined for each node $v_z$ to capture the properties of first-order and second-order neighbours, respectively. In these matrices, rows represent k-shell index and columns represent degree. 
The maximum k-shell and degree in the network are denoted by $s$ and $d$, respectively, hence each of these matrices has a size of $s \times d$. The value of each element of the matrix $A^{(r)}(z)$ is given by Eq.~(\ref{Eq.1}).
\begin{equation}
A_{ij}^{(r)}(z)=\frac{z_{ij}^{(r)}}{\sum_{i=1}^{s}{\sum_{j=1}^{d}{z_{ij}^{(r)}}}} % 
\label{Eq.1}
\end{equation}
where $r$ is to differentiate between first-order neighbours ($r=1$) and second-order neighbours ($r=2$) of node $v_z$.
In the equation, $z_{ij}^{(1)}$ is the number of the first-order neighbours of node $v_z$ which have k-shell=$i$ and degree=$j$. The value of $z_{ij}^{(1)}$ is calculated using Eq.~(\ref{Eq.2}).
\begin{equation}
z_{ij}^{(1)}=|\{ v_w \mbox{  where  } ks_w=i, d_w=j,  v_w\in N_z \}|
\label{Eq.2}
\end{equation}
 $z_{ij}^{(2)}$ is also the number of the second-order neighbours of node $v_z$ which have k-shell=$i$ and degree=$j$. The value of $z_{ij}^{(2)}$ is calculated using Eq.~(\ref{Eq.3}).
 \begin{equation}
z_{ij}^{(2)}=|\{ v_x \mbox{  where  } ks_x=i,d_x=j,v_x\in N_w,v_w\in N_z \}|
\label{Eq.3}
\end{equation}

Then, the covariance value of each matrix $A^{(1)}(z)$ and $A^{(2)}(z)$ is calculated; this is a value between $-1$ and $1$. If there is a direct relation between degree and k-shell in the neighbourhood of node $v_z$ the covariance takes a value close to 1. That is to say, if the value of degree increases by increasing k-shell in the neighbourhood of the node, the matrices have larger values of covariance, and the node is accordingly more influential. To calculate the covariance of matrix $A^{(r)}(z)$, Eq.~(\ref{Eq.4}) is used.
\begin{equation}
Cov(A^{(r)}(z))=E(A^{(r)}(z))-E(S^{(r)}(z))\cdot E(D^{(r)}(z))
\label{Eq.4}
\end{equation}
where $S^{(r)}(z)$ and $D^{(r)}(z)$ are marginal distributions of matrix $A^{(r)}(z)$ and are calculated using Eqs.~(\ref{Eq.5}) and (\ref{Eq.6}). In these equations $A_{ij}^{(r)}(z)$ is the entry of matrix $A^{(r)}(z)$ in row $i$ and column $j$.
\begin{equation}
S_i^{(r)}(z)=\sum_{j=1}^{d}A_{ij}^{(r)}\;\; i=1,...,s
\label{Eq.5}
\end{equation}
\begin{equation}
D_i^{(r)}(z)=\sum_{j=1}^{s}A_{ji}^{(r)}\;\; i=1,...,d
\label{Eq.6}
\end{equation}
In Eq.~(\ref{Eq.4}), $E(S^{(r)}(z))$ and $E(D^{(r)}(z))$ are the expectation of the marginal distribution $S^{(r)}(z)$ and $D^{(r)}(z)$, respectively.
 $E(A^{(r)}(z))$ is also the expectation of matrix $A^{(r)}(z)$. Their values are calculated using Eqs.~(\ref{Eq.7}) to (\ref{Eq.9}).
 \begin{equation}
E(S^{(r)}(z))=\sum_{i=1}^{s}\frac{i}{s}\,S_i^{(r)}(z)
\label{Eq.7}
\end{equation}
 \begin{equation}
E(D^{(r)}(z))=\sum_{i=1}^{d}\frac{i}{d}\,D_i^{(r)}(z)
\label{Eq.8}
\end{equation}
 \begin{equation}
E(A^{(r)}(z))=\sum_{i=1}^{s}\sum_{j=1}^{d}\frac{i}{s}\,\frac{j}{d}\,A_{ij}^{(r)}(z)
\label{Eq.9}
\end{equation}

Then, the SDC value for node $v_z$ is calculated using Eq.~(\ref{Eq.10}), based on the values obtained for $Cov(A^{(1)}(z))$ and $Cov(A^{(2)}(z))$ from Eq.~(\ref{Eq.4}).
\begin{equation}
\begin{multlined}
SDC(v_z)=\frac{d_z}{d}\,(2+Cov(A^{(1)}(z)))+\\ \frac{d_z^{(2)}}{d^{(2)}}\,(2+Cov(A^{(2)}(z)))
\end{multlined}
\label{Eq.10}
\end{equation}
In this equation, $d_z^{(2)}=\sum_{v_w\in N_z} d_w$ is the second-order degree of node $v_z$, and $d^{(2)}$ is the maximum second-order degree in the network. 
% RIZOS - why do you do this +2?
% Reply - because , here, the value of covariance shows the help of the neighbours to spread a message. even a node has neighbours with negative or zero covariance, they do not have negative or neutral role in spreading influence of the node. I mean having such neighbours is better than not having them. so by adding 2, I try to change the value of variance between -1 and +1, to a value between +1 and +3.    
%Given that the covariance obtained in Eq.~(\ref{Eq.4}) has values between $-1$ and $+1$ 
%but from SDC's point of view we are more interested in nodes with high positive covariance, 
%the value of 2 is added in Eq.~(\ref{Eq.10}) to create positive only values for SDC. 

Algorithm~\ref{Algo:1} shows pseudo-code of the proposed method for calculating the SDC centrality of node $v_z$.

\begin{algorithm}[!ht]
\SetAlgoLined
\KwData{Graph $G(V,E)$ and node $v_z$}
\KwResult{The value of SDC centrality of node $v_z$}
\lstset{numbers=left, numberstyle=\tiny, stepnumber=1, numbersep=5pt}
          set $r$=1 \;
          calculate matrix $A^{(r)}(z)$ using Eq.~(\ref{Eq.1})\;
          calculate the marginal distributions $S^{(r)}(z)$ and $D^{(r)}(z)$ using Eqs.~(\ref{Eq.5}) and (\ref{Eq.6}),   respectively\; 
          calculate $E(S^{(r)}(z))$, $E(D^{(r)}(z))$ and  $E(A^{(r)}(z))$, using Eqs. (\ref{Eq.7}), (\ref{Eq.8}) and (\ref{Eq.9}), respectively\;
          calculate $Cov(A^{(r)}(z))$ using Eq.~(\ref{Eq.4})\;
          set $r$=2 and repeat steps 2 to 5\;
          calculate SDC($v_z$) using Eq.~(\ref{Eq.10})
 \caption{SDC calculation for $v_z$}
 \label{Algo:1}
\end{algorithm}

Although SDC centrality can be used to determine the spreading capability of nodes too, considering the centrality of the neighbours increases the accuracy of this determination~\cite{neighborhoodRule}. %Taking the centrality of neighbours up, to second-order neighbours, increases the accuracy of the process of determining nodes' influence and ranking nodes~\cite{neighborhoodRule}. 
To achieve this, we cumulatively sum up the centrality of neighbours (indirectly up to second-order neighbours) using Eqs.~(\ref{Eq.11}) and (\ref{Eq.12}), 
%As stated in Eqs.~(\ref{Eq.11}) and (\ref{Eq.12}), 
where the Covariance Centrality (CVC) and the Extended Covariance Centrality (ECVC) for node $v_z$ are defined based on the SDC of its neighbours.
 \begin{equation}
CVC(v_z)=\sum_{V_w\in N_i}SDC(v_w)
\label{Eq.11}
\end{equation}
 \begin{equation}
ECVC(v_z)=\sum_{V_w\in N_i}CVC(v_w)
\label{Eq.12}
\end{equation}
We calculate the ECVC of every node of the network to determine a node's spreading ability and produce a ranking list that ranks nodes in descending order of their influence. 
Algorithm~\ref{Algo:2} shows the pseudo-code of the proposed method to rank the nodes.  

\begin{algorithm}[!ht]
\SetAlgoLined
\KwData{Graph $G(V,E)$}
\KwResult{ Ranking list (list of nodes in descending order of influence)}
\lstset{numbers=left, numberstyle=\tiny, stepnumber=1, numbersep=5pt}
      calculate  SDC($v_z$) for each $v_z\in V$ using Algorithm~\ref{Algo:1}\;
      calculate CVC($v_z$) for each $v_z\in V$ using Eq.~(\ref{Eq.11})\;
      calculate ECVC($v_z$) for each $v_z\in V$ using Eq.~(\ref{Eq.12})\;
      rank the nodes in descending order according to their ECVC value.
 \caption{ECVC ranking pseudo-code}
 \label{Algo:2}
\end{algorithm}

The time complexity of Algorithm~\ref{Algo:1} can be analysed as follows. The calculation of the $A^{(1)}(z)$ matrix for each node $v_z$ needs $O( \langle d \rangle)$, where $ \langle d \rangle$ is the average degree of nodes. In line 3 of the algorithm, it takes $O(s)$ and $O(d)$ to calculate $S^{(1)}$ and $D^{(1)}$, respectively. The calculation of Eqs.~(\ref{Eq.7}), (\ref{Eq.8}) and (\ref{Eq.9}) in line 4 of the algorithm needs $O(s)$, $O(d)$, and $O(s\times d)$, respectively. Line 5 is $O(1)$.  
These steps must be repeated for $r=2$. The complexity is the same as $r=1$ with the exception of line 2 which is $O(\langle d^{(2)}\rangle)$; $\langle d^{(2)}\rangle$ is the average value of second-order degree of nodes. Thus, the time complexity of Algorithm~\ref{Algo:1} is $O(\langle d \rangle + \langle d^{(2)}\rangle+2\times(2 s+2 d+s\times d+1)) \in O(\langle d^{(2)}\rangle)$.  
In Algorithm~\ref{Algo:2}, the SDC centrality of every node must be calculated using Algorithm~\ref{Algo:1}, so it has time order $O(|V| \times \langle d^{(2)}\rangle)$. The calculation of CVC and ECVC have complexity $O(\langle d \rangle)$ and $O(\langle d^{(2)}\rangle)$, respectively. The time complexity of the k-shell algorithm to determine the shells is $O(|E|)$~\cite{Cnc+}. Thus, the total complexity of the proposed method is $O(|E|+|V| \times \langle d^{(2)}\rangle +\langle d \rangle+\langle d^{(2)}\rangle) \in O(|E| + |V| \times \langle d^{(2)}\rangle)$.

\begin{figure}[ht!]
\begin{center}
\includegraphics[scale=0.32]{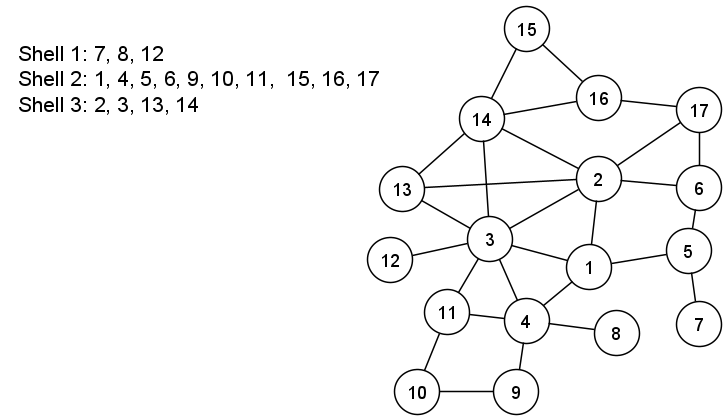}
\caption{An example network}
\label{Fig_Example}
\end{center}
\end{figure}

% ****************************************************
\subsection{An example}
As an example, to understand the proposed method, consider Fig.~\ref{Fig_Example}. In this graph the maximum shell, $s$, and maximum degree, $d$ are 3 and 7, respectively. The nodes located in each shell are shown in the figure. We will show how to calculate the value of ECVC centrality for node 2, $v_2$.

Node 2, $v_2$, has 6 neighbours, i.e $N_2=\{v_1,v_3,v_6,v_{13},v_{14},v_{17}\}$ and $d_2=6$. The matrix $A^{(1)}(2)$ for this node is calculated as follows.

\[
  A^{(1)}(2)=
  \begin{pmatrix}
\frac{0}{6}&\frac{0}{6}&\frac{0}{6}&\frac{0}{6}&\frac{0}{6}&\frac{0}{6}&\frac{0}{6}\\
&&&&&&\\
\frac{0}{6}&\frac{0}{6}&\frac{2}{6}&\frac{1}{6}&\frac{0}{6}&\frac{0}{6}&\frac{0}{6}\\
&&&&&&\\
\frac{0}{6}&\frac{0}{6}&\frac{1}{6}&\frac{0}{6}&\frac{1}{6}&\frac{0}{6}&\frac{1}{6}\\
\end{pmatrix}
\]

The value of the element in row $i$ and column $j$ of the matrix indicates what is the proportion of neighbours with k-shell $i$ and degree $j$ that node 2 has. Consider for instance the element in row 2 and column 3 of the matrix. The value of $\frac{2}{6}$ means that out of a total of six neighbours $v_2$ has two neighbours with k-shell 2 and degree 3; indeed, these nodes are $v_6$ and $v_{17}$. 

We use Eq.~(\ref{Eq.5}) to calculate the vector $S^{(1)}(2)$. For example, the value of the third element of $S^{(1)}(2)$ is calculated as $S_3^{(1)}(2)=A_{3,1}^{(1)}(2)+A_{3,2}^{(1)}(2)+A_{3,3}^{(1)}(2)+A_{3,4}^{(1)}(2)+A_{3,5}^{(1)}(2)+A_{3,6}^{(1)}(2)+A_{3,7}^{(1)}(2)=\frac{0}{6}+\frac{0}{6}+\frac{1}{6}+\frac{0}{6}+\frac{1}{6}+\frac{0}{6}+\frac{1}{6}=\frac{3}{6}$. The value of the other entries of $S^{(1)}(2)$ is calculated similarly. Hence, $S^{(1)}(2)$ is given by  
\[
S^{(1)}(2)=
\begin{pmatrix}
\frac{0}{6}&\frac{3}{6}&\frac{3}{6}\\
\end{pmatrix}
\]

We use Eq.~(\ref{Eq.6}) to calculate the vector $D^{(1)}(2)$. For example, the value of the fourth element of $D^{(1)}(2)$ is calculated as $D_4^{(1)}(2)=A_{1,4}^{(1)}(2)+A_{2,4}^{(1)}(2)+A_{3,4}^{(1)}(2)=\frac{0}{6}+\frac{1}{6}+\frac{0}{6}=\frac{1}{6}$. Overall, $D^{(1)}(2)$ is given by
\[D^{(1)}(2)=
\begin{pmatrix}
\frac{0}{6}&\frac{0}{6}&\frac{3}{6}&\frac{1}{6}&\frac{1}{6}&\frac{0}{6}&\frac{1}{6}\\
\end{pmatrix}
\]
The expectation of the marginal distribution $S^{(1)}(2)$ and $D^{(1)}(2)$ and matrix $A^{(1)}(2)$ is calculated using Eqs.~(\ref{Eq.7},\ref{Eq.8},\ref{Eq.9}) as
\[
\begin{split}
E(S^{(1)}(2))=%\frac{1}{3}\cdot S^{(1)}[1]+\frac{2}{3}\cdot S^{(1)}[2]+\frac{3}{3}\cdot S^{(1)}[3]\\=
\frac{1}{3}\cdot \frac{0}{6}+\frac{2}{3}\cdot \frac{3}{6}+\frac{3}{3}\cdot \frac{3}{6}\cong 0.83
\end{split}
\]
\[
\begin{split}
E(D^{(1)}(2))\!=\!
%\frac{1}{7}\cdot D^{(1)}[1]+\frac{2}{7}\cdot D^{(1)}[2]+\frac{3}{7}\cdot D^{(1)}[3]+\\\frac{4}{7}\cdot %D^{(1)}[4]+\frac{5}{7}\cdot D^{(1)}[5]+\frac{6}{7}\cdot D^{(1)}[6]+\frac{7}{7}\cdot D^{(1)}[7]=\\
\frac{1}{7}\!\cdot\!\frac{0}{6}\!+\!\frac{2}{7}\!\cdot\!\frac{0}{6}\!+\!\frac{3}{7}\!\cdot\!\frac{3}{6}\!+\!\frac{4}{7}\!\cdot\!\frac{1}{6}\!+\!\frac{5}{7}\!\cdot\!\frac{1}{6}\!+\!\frac{6}{7}\!\cdot\!\frac{0}{6}\!+\!\frac{7}{7}\!\cdot\!\frac{1}{6} \cong 0.60
\end{split}
\]
\[
\begin{split}
E(A^{(1)}(2))=\frac{1}{3}\cdot \frac{1}{7}\cdot A_{1,1}^{(1)}(2)+ \frac{1}{3}\cdot \frac{2}{7}\cdot A_{1,2}^{(1)}(2)+\frac{1}{3}\cdot \frac{3}{7}\cdot A_{1,3}^{(1)}(2)+\dots = \\ = \frac{1}{3}\cdot \frac{1}{7}\cdot \frac{0}{6}+ \frac{1}{3}\cdot \frac{2}{7}\cdot \frac{0}{6}+\frac{1}{3}\cdot \frac{3}{7}\cdot \frac{0}{6}+\dots\cong 0.52
\end{split}
\]
The value of $Cov(A^{(1)}(2))$ is calculated using Eq.~(\ref{Eq.4}) as 
\[
\begin{split}
Cov(A^{(1)}(2))=E(A^{(1)}(2))-E(S^{(1)}(2))\cdot E(D^{(1)}(2)) \cong 0.52-0.83\cdot 0.60\cong 0.02
\end{split}
\]

The same process is repeated for the second-order neighbours of $v_2$ and the value of $Cov(A^{(2)}(2))$ obtained is $Cov(A^{(2)}(2))\cong 0.04$. The value of the SDC centrality of node 2, $v_2$, is calculated using Eq.~(\ref{Eq.10}) as 
\[
\begin{split}
SDC(2) = & \frac{6}{7}\cdot (2+Cov(A^{(1)}(2))) + \frac{25}{27}\cdot (2+Cov(A^{(2)}(2))) \cong \\ 
\cong & \frac{6}{7}\cdot (2+0.02)+ \frac{25}{27}\cdot (2+0.04)\cong 3.62 
\end{split}
\]
The value of SDC centrality is similarly calculated for all nodes, and then the CVC centrality of node 2, $v_2$, is obtained using Eq.~(\ref{Eq.11}) as
\[
\begin{split}
CVC(2)=SDC(1)+SDC(3)+SDC(6)+SDC(13)+SDC(14)+SDC(17)\cong \\\cong 2.75+4.08+1.78+2.21+3.03+1.77 \cong 15.61
\end{split}
\]
After the calculation of the CVC centrality for all nodes, the value of ECVC centrality of node 2, $v_2$, is obtained using Eq.~(\ref{Eq.12}) as 
\[
\begin{split}
ECVC(2)=CVC(1)+CVC(3)+CVC(6)+CVC(13)+CVC(14)+CVC(17)\cong \\\cong 11.91+17.10+6.86+10.72+12.70+7.01 \cong 66.31.
\end{split}
\]

% ***********************************************************************************

\section{Experimental Results}
\label{Section:5}
This section evaluates the performance of the proposed method, ECVC (as well as the measures that compose it, SDC and CVC), in comparison with another eleven centrality measures that can be used to identify and rank influential nodes. The eleven methods are: k-shell (KS)~\cite{kshell}, degree (D)~\cite{degree}, mixed degree decomposition (MDD)~\cite{MDD}, extended neighbourhood coreness (Cnc+)~\cite{Cnc+}, k-shell iteration factor (KS-IF)~\cite{kshell-if}, degree and importance of links (DIL)~\cite{DIL}, Mixed Core, Semi-local Degree and Weighted Entropy (MCDE)~\cite{MCDE}, link significance (LS)~\cite{LS}, hierarchical k-shell (HKS)~\cite{HKS}, diversity-strength ranking (DSR)~\cite{EDSR}, and weighted k-shell degree (WKD) \cite{KSD}.

A set of {twelve} real and two artificial (LFR-200 and LFR-1000) networks are used, as shown in Table~\ref{Tble_Stats}. The real world networks include: Karate club(KRT)~\cite{DatasetKRT}, Dolphins (DLN)~\cite{DatasetDLN}, Jazz musician (JZM)~\cite{DatasetJZM}, Netsciense (NTS)~\cite{DataSetFFD}, Facebook Food (FFD)~\cite{DatasetELS}, Euroroad (EUR)~\cite{DatasetEUR}, Euroroad (EUR)~\cite{DatasetEUR}, Chicago (CHG)~\cite{DatasetCHG}, Chicago (CHG)~\cite{DatasetCHG}, Hamsterster (HMS)~\cite{DatasetHAM}, Ego-Facebook (FCB)~\cite{FaceBook}, US Power grid (UPG)~\cite{DatasetUPG}, LastFM (LFM)~\cite{LastFM} and Pretty Good Privacy (PGP)~\cite{DatasetPGP}.     
{For each network, the table shows: number of nodes $(|V|)$, number of edges $(|E|)$, maximum degree $(MD)$,  average degree of nodes $(\langle d\rangle)$, average second-order degree of nodes $(\langle d^{(2)}\rangle)$ and average clustering coefficient $(CC)$.} 
The Lancichinetti–Fortunato–Radicchi (LFR) benchmark~\cite{LFR} is used to generate the artificial networks. LFR is able to generate an artificial network with a set of defined parameters, such as, number of nodes $(|V|)$, average degree of nodes $(\langle d\rangle)$, mixing parameter of the community structure $(\mu)$, and power-law of the degree distribution $(\gamma)$.  These parameters are set as $|V|=200$, $\langle d \rangle=5$, $\gamma=2$, and $\mu=0.2$ to generate LFR-200, and they are set as $|V|=1000$, $\langle d\rangle=10$, $\gamma=2$, and $\mu=0.2$ to generate LFR-1000.

Two different criteria are used to evaluate the ranking list produced by the different methods for different networks. The first criterion is discrimination of the nodes with different spreading ability, which is the capability to differentiate between nodes according to their influence. The second criterion is accuracy (or correctness), which assesses the ranking list and the nodes in terms of their spreading ability using a standard model that simulates the spreading process.
An experiment is also dedicated to evaluating the average running time of all the methods.

%*******************************************************
 
 \begin{figure*}[!tb]
 \center{
 
 \subfloat[FCB]{
   \centering
       \includegraphics[width=0.5\textwidth]{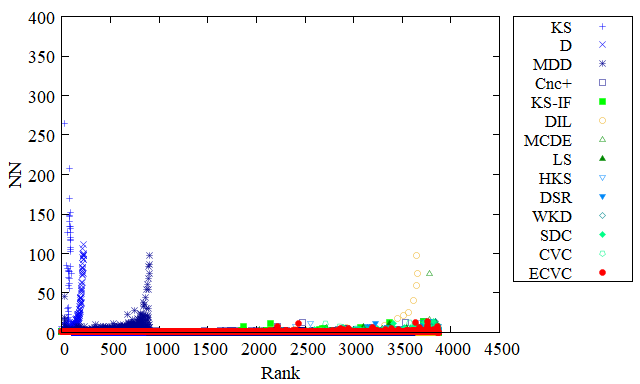}\label{NN_FCB}
    }
 \subfloat[UPG]{
   \centering
       \includegraphics[width=0.5\textwidth]{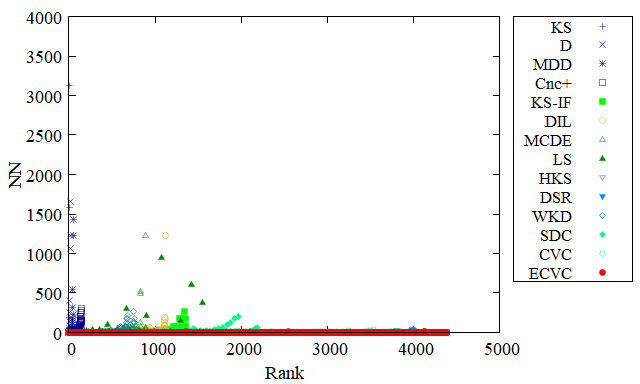}\label{NNUPG}
    }

 \subfloat[LFM]{
   \centering
       \includegraphics[width=0.5\textwidth]{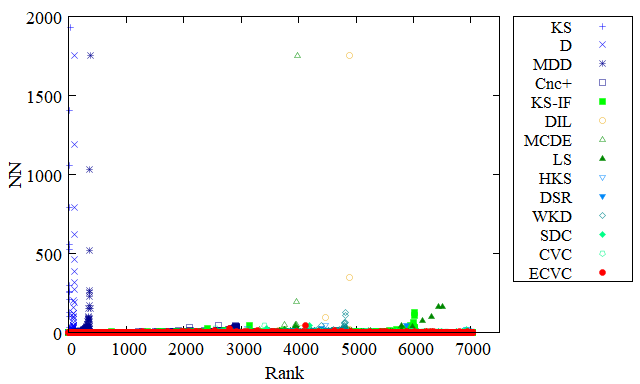}\label{NNLFM}
    }
        \subfloat[PGP]{
   \centering
       \includegraphics[width=0.5\textwidth]{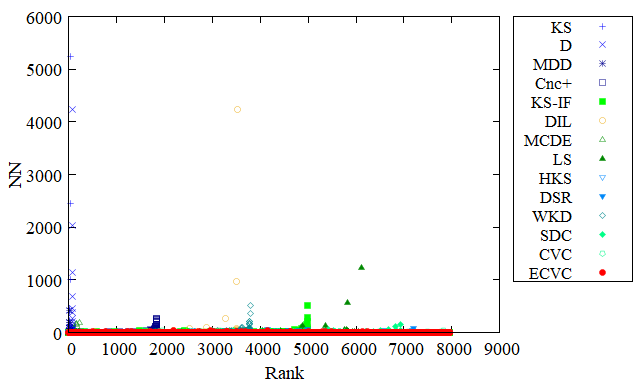}\label{NNPGP}
    }
       \caption{The number of nodes ($NN$) in different ranks of $R$ by various methods in FCB, UPG, LFM and PGP networks}
       \label{Fig_NN}
       }
 \end{figure*}
 
\subsection{Discrimination}
The discrimination of the ranking list $R$ produced by different methods is investigated in this section. A distinct metric (DM) function~\cite{FunctionDM} is applied for this purpose. The value of the DM function for the list $R$ is calculated using Eq.~(\ref{Eq.13}). The value of DM is in the interval $(0,1]$; a larger value shows better discrimination in $R$.
\begin{equation}
    DM(R)=\frac{\mbox{number of distinct ranks}}{|V|}
    \label{Eq.13}
\end{equation}
A rank refers to a group of nodes with the same influence value (as computed by some centrality measure). Clearly, if all nodes have a different value maximum discrimination is achieved (DM=1)~\cite{FunctionDM}.

The results obtained are shown in Table~\ref{Tble_DM}. ECVC considers the relation between degree and k-shell for nodes' neighbours to discriminate the spreading ability of nodes which are located in the same shell. As seen from the table, ECVC has overall better distinction in comparison to other methods. 
 {ECVC significantly outperforms all  methods in larger networks. In three relatively small networks, such as NTS, L1000 and FCB, DSR performs better than ECVC but this is by a small margin. Overall ECVC shows better properties than other measures.}

For a better evaluation of the discrimination ability of the proposed method, the number of nodes (NN) in different ranks of the ranking list $R$ is investigated in the next experiment of this section.  Fig.~\ref{Fig_NN} shows the obtained results of the experiment using  {the four largest networks, FCB, UPG, LFM, PGP, which have different features}. It can be seen from Fig.~\ref{Fig_NN} that ECVC assigns a small number of nodes to the same rank. This means that nodes are ranked in more ranks by ECVC as opposed to other methods, something that leads to better discrimination, especially in large networks.

%****************************************************

\subsection{Accuracy}

In this section, the accuracy of the ranking list $R$ produced by different methods is evaluated using three sets of experiments. To do so, a susceptible-infected-recovered (SIR) model~\cite{SIR1,SIR2} is used to simulate the spreading process in the real world and determine the ground-truth influence of nodes. The real influence ranking list $(\sigma)$ is determined based on the ground-truth influence of nodes. %Then, the correlation between $\sigma$ and the ranking list $R$, as produced by different methods, is calculated. 

SIR is one of the epidemic diffusion models which has been widely used in research~\cite{Cnc+,CN,LS,EDSR} to simulate spreading process and generate a real influence ranking list. In this model, each node can be in one of susceptible, infected, or recovered states. Node $v_i$ is initially considered as an infected node and all other nodes are considered as susceptible. In each timestamp, each infected node is changed to recovered state after its attempt to infect each susceptible neighbour with infection probability $\beta$. This process continues until there is not any infected node in the network. The number of recovered nodes is considered as ground-truth influence of node $v_i$. This epidemic process is repeated for every node  $(i=1,\dots,|V|)$ and the real influence ranking list $(\sigma)$ is accordingly determined. In our experiments, in line with other studies~\cite{Cnc+,KSD,H-index,EDSR}, to increase the accuracy in determining $\sigma$, the SIR model is repeated 1000 times for each node and the average value of the number of recovered nodes is considered as influence. The value of $\beta$ must be chosen so that it is slightly greater than the infection threshold $\beta_{th}=\frac{\langle d \rangle}{\langle d^{(2)}\rangle}$~\cite{Cnc+}, where $\langle d \rangle$ and $\langle d^{(2)}\rangle$ are the average of first- and second-order degrees of the network nodes, respectively. 
The values of $\beta_{th}$ and $\beta$ considered for each network are given in Table~\ref{Tble_Beta}.

The Kendall tau correlation coefficient $(\tau)$~\cite{Kendal} is employed to calculate the correlation between the ranking lists $R$ and $\sigma$ for each different method. A higher value of correlation between $\sigma$ and $R$ shows better accuracy of the ranking list $R$. For this purpose, concordant and discordant pair sets between $R$ and $\sigma$ are determined. Suppose that $(\sigma_1,R_1),(\sigma_2,R_2),\dots ,(\sigma_n,R_n),$ is a set of pair ranks in $R$ and $\sigma$. If $(\sigma_i>\sigma_j \mbox{  and  } R_i>R_j)$ or $(\sigma_i<\sigma_j \mbox{  and  } R_i<R_j)$ then $(\sigma_i,R_i)$ and $(\sigma_j,R_j)$ are considered as concordant pairs. Conversely, they are considered as discordant pairs if  $(\sigma_i>\sigma_j \mbox{  and  } R_i<R_j)$ or $(\sigma_i<\sigma_j \mbox{  and  } R_i>R_j)$. The Kendal tau correlation coefficient, $\tau(\sigma ,R)$, is calculated with the help of Eq.~(\ref{Eq.14}) and it is a real number in [$-1, +1$]. A value of zero indicates no correlation between the lists, whereas values closer to 1 show high positive correlation and values closer to $-1$ show high negative correlation. 
\begin{equation}
    \tau(\sigma ,R)=\frac {n_c-n_d}{\frac{1}{2}(n)(n-1)} 
    \label{Eq.14}
\end{equation}

\begin{figure*}[!tb]
 \center{
 
 \subfloat[FFD]{
   \centering
       \includegraphics[width=0.5\textwidth]{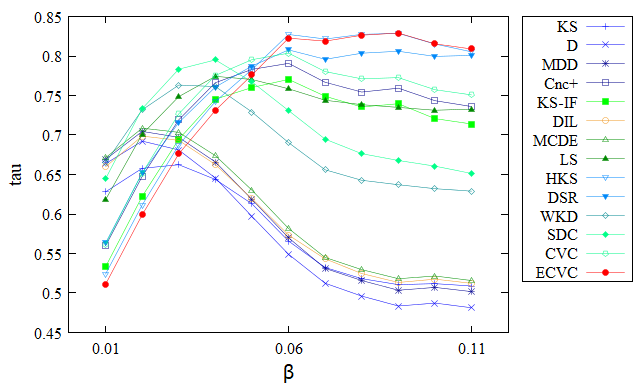}\label{TFFD}
    }
 \subfloat[NTS]{
   \centering
       \includegraphics[width=0.5\textwidth]{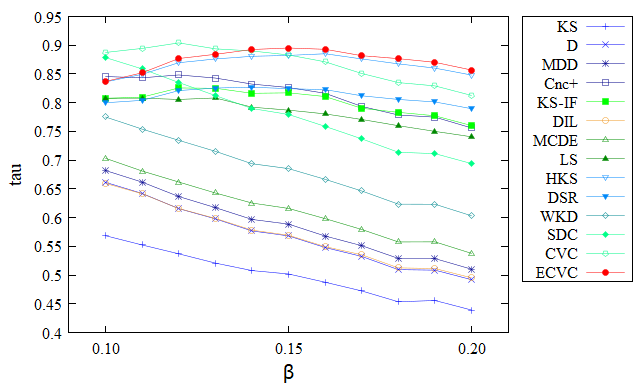}\label{TNTS}
    }
    
 \subfloat[CHG]{
   \centering
       \includegraphics[width=0.5\textwidth]{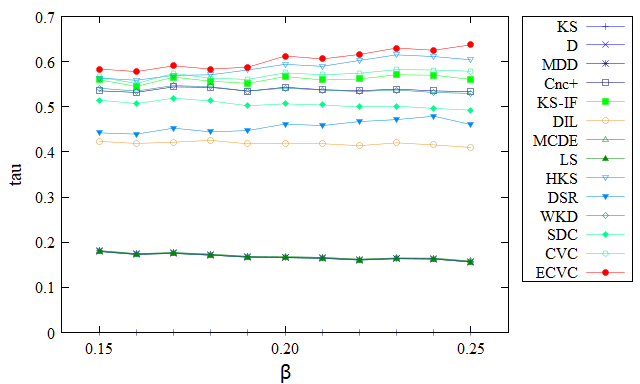}\label{TCHG}
    }
 \subfloat[EUR]{
   \centering
       \includegraphics[width=0.5\textwidth]{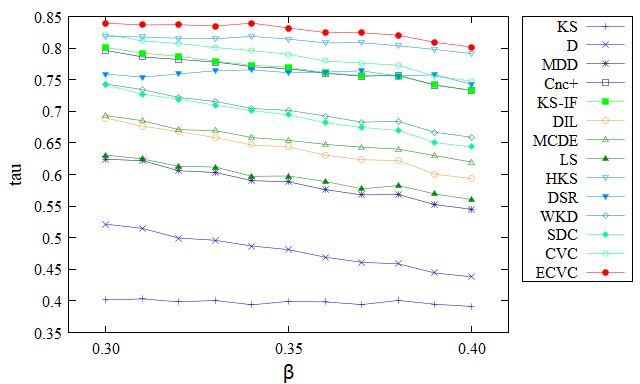}\label{TEUR}
    }
\caption{The effect of varying the value of the infection probability on the accuracy of different methods using the FFD, NTS, CHG, and EUR networks}
\label{Fig_Beta}
}
 \end{figure*}

\begin{figure*}[!tb]
 \center{
 
 \subfloat[FCB]{
   \centering
       \includegraphics[width=0.5\textwidth]{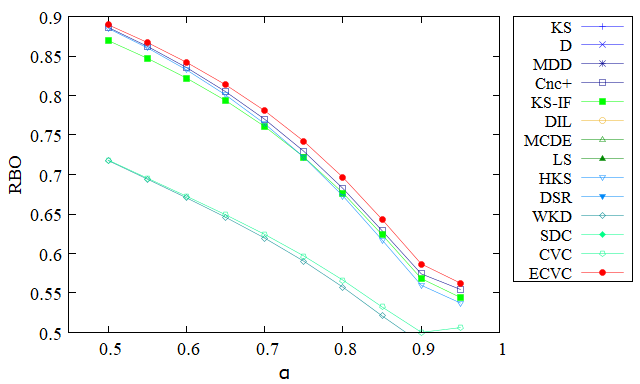}\label{RFCB}
    }
 \subfloat[UPG]{
   \centering
       \includegraphics[width=0.5\textwidth]{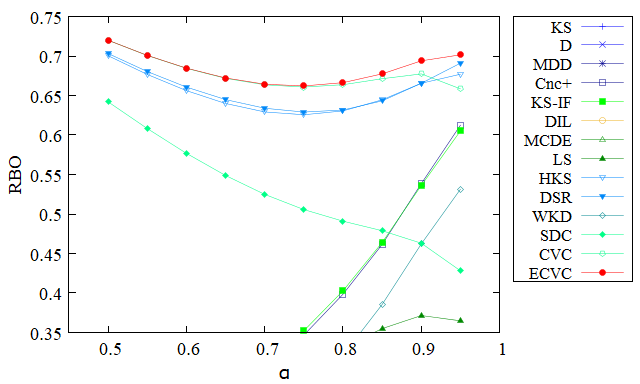}\label{RUPG}
    }

 \subfloat[LFM]{
   \centering
       \includegraphics[width=0.5\textwidth]{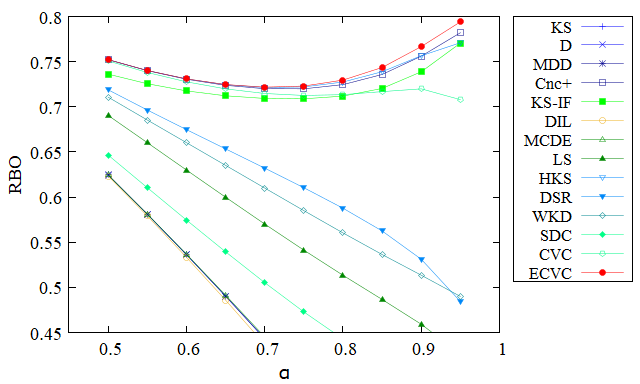}\label{RLFM}
    }
    \subfloat[PGP]{
   \centering
       \includegraphics[width=0.5\textwidth]{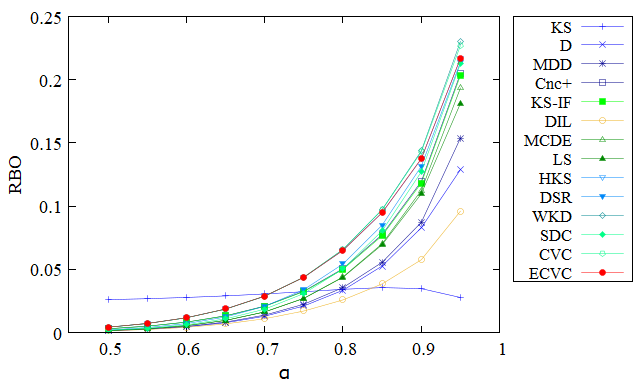}\label{RPGP}
    }
       \caption{The value of RBO for different methods using the FCB, UPG, LFM and PGP networks}
       \label{Fig_RBO}
}
 \end{figure*}

The first experiment compares  {ECVC, SDC, CVC and the eleven different methods from the literature}, which are used to determine a ranking list $R$; then, the Kendall tau correlation coefficient $(\tau)$ between $R$ and $\sigma$ is calculated. Table~\ref{Tble_Tau} shows the value of $\tau$ for all networks and methods.
It can be seen from Table~\ref{Tble_Tau} that ECVC outperforms other methods most of the time. %HKS has the closest results to it.
HKS has higher accuracy than ECVC for the KRT, FFD and UPG networks and equal accuracy for L1000 %LS, for the ELS network, and
 {but for all other networks ECVC has the highest accuracy. This is because degree and k-shell of neighbours are concurrently taken into account: nodes whose neighbours have high degree and appropriate topological location are considered as influential nodes by ECVC. Focusing on the three methods proposed in this paper, ECVC, SDC, and CVC, it can be seen that ECVC outperforms the other two methods, except in the case of the KRT and JZM networks. This is because these two networks are small, yet the average second-order degree is larger than the number of nodes, which suggests that ECVC may overestimate structural information.}

In the second experiment, we assess the effect of varying the infection probability on the accuracy of ECVC. 
For this purpose, this value is varied %$\beta_{th}$. %from $\beta_{th}-0.14$ to $\beta_{th}+0.05$ in steps of 0.01 (a total of 20 different values). 
%$\beta+\lambda $ where $\lambda$ varied 
from $\beta-0.05$ to $\beta+0.05$ in steps of 0.01 (a total of 11 different values).
The ranking list $\sigma$ is determined using the SIR model. The Kendall tau correlation coefficient between $\sigma$ and ranking list $R$ is then calculated. The results obtained using four representative networks, % {with different value of $\beta_{th}$, 
FFD, NTS, CHG, and EUR, are shown in Fig.~\ref{Fig_Beta}. 
 {As the infection probability increases, the message can spread to farther distances. Then, the first and second-order neighbours play a more important role in the spreading process. When the value of $\beta$ is very small, second-order neighbours become less important in the spreading process. Thus, for small values of $\beta$, methods that rely on first-order neighbours to determine the spreading ability of nodes are reasonably accurate, something that can be observed in Fig.~\ref{Fig_Beta}a. As the value of $\beta$ increases, the second-order neighbours play a more important role in the spreading process. In this case, methods relying on first-order neighbours have lower accuracy whereas the accuracy of methods that take additional information in the neighbourhood into account, increases. As seen in Fig.~\ref{Fig_Beta}, ECVC has the highest accuracy and outperforms all other methods in all cases, except for small values of $\beta$ in Fig.~\ref{Fig_Beta}a, for the reasons already discussed.
}
%although Cnc+ has the closest results to ECVC in the KRT network, in other networks HKS is the second best method after ECVC.

% RIZOS - explain what top-rank is and what low-rank is
% Reply - the nodes which are identified as more influential and assigned to the top-ranks (nodes in rank 1 are the most influential). Accuracy in top-ranks is more important because we consider the nodes in top-ranks as more influential and use them for different application like influence maximization and rumor block and so on.

The third experiment is dedicated to the evaluation of the accuracy of the proposed method with respect to top-ranks. This is of interest as most influential nodes belong to top-ranks in the ranking list (recall the definition of rank in the description of Eq.~(\ref{Eq.13}). For this purpose, the rank-biased overlap (RBO) function~\cite{FunctionRBO} is employed to assess the overlap of top-ranks in the real influence ranking list, $\sigma$ with top-ranks in the ranking list, $R$, produced by each method. The value of RBO is calculated as in Eq.~(\ref{Eq.15}). % to determine the accuracy of top-f ranks of $R$.
 \begin{equation}
RBO(\sigma , R , \alpha)=(1-\alpha) \sum _{f=1}^{n}\alpha ^{f-1} A(\sigma , R , f)
\label{Eq.15}
\end{equation}
where $n$ is the maximum number of distinct ranks in the ranking lists $\sigma$ and $R$. % and $f$ iterates over all ranks. 
The parameter
% RIZOS - check what values of \alpha make sense
% Reply - A lower value of alpha gives higher weight to top-ranks and accuracy in top-ranks plays more significant role in determination of RBO.
$\alpha$ can take values between 0 and 1. A lower value of $\alpha$ gives higher importance to top-ranks in the ranking list $R$. $A (\sigma, R, f)$ is the amount of overlap between the first $f$ top-ranks of the lists and is calculated using Eq.~(\ref{Eq.16}). In this equation, $\sigma(f)$ and $R(f)$ are the set of nodes in the $f$ top-ranks in lists $\sigma$ and $R$, respectively.
\begin{equation}
A(\sigma , R , f)=\frac {|\sigma (f) \cap R(f)|}{|\sigma (f) \cup R(f)|} 
\label{Eq.16}
\end{equation}
The value of RBO function is in [0, 1]. A higher value of this function shows more overlap between the top-ranks of the lists, and, accordingly, a higher accuracy of the top-ranks of list $R$. The results for this experiment, using the four largest networks FCB, UPG, LFM and PGP, are shown in Fig.~\ref{Fig_RBO}. 
 {A lower value of $\alpha$ gives greater weight to the top ranks in the ranking list. This means that 
a downward trend in Fig.~\ref{Fig_RBO} suggests that the ranking list has greater accuracy for top ranks but its accuracy decreases going toward the lower ranks. Conversely, an upward trend suggests less accuracy in top ranks but the accuracy increases as we give more weight to the lower ranks. As can be seen from Fig.~\ref{Fig_RBO}, ECVC outperforms all other methods consistently with the only exception being the PGP network. This may be because, in this network, nodes are ranked in a relatively small number of ranks with many nodes in each rank, something that may potentially decrease the estimated overlap according to Eq.~\ref{Eq.16}.
%. In UPG and LFM, ECVC has great accuracy in both top and low ranks; although ECVC is followed closely by CVC and Cnc+ in UPG and LFM, respectively, it outperforms all other methods. In PGP, all methods have lower accuracy in top ranks and their accuracy increases in lower ranks. In this network, KS has the greatest accuracy in top ranks; WKD slightly outperforms ECVC in PGP but ECVC has significant greater accuracy than WKD in other networks. 
%In general, ECVC is more accurate in identifying top-ranks for larger networks. This is because influential nodes are located closer to the core of the network and have a higher degree. The elaborate consideration of the relationship between these features by ECVC leads to its higher accuracy compared to other methods.
}

%****************************************

\subsection{Running Time}

\begin{figure}[ht!]
\begin{center}
\includegraphics[scale=0.5]{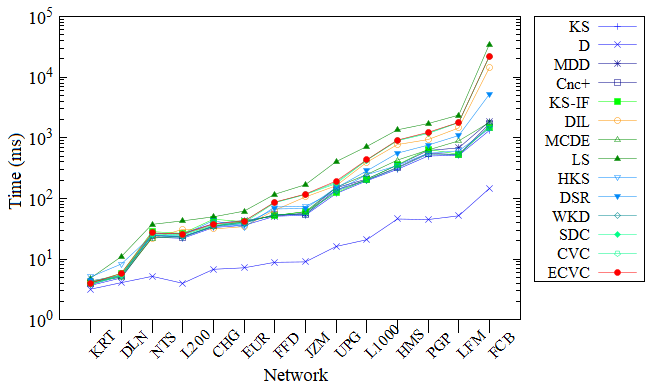}
\caption{Running time (in seconds) of the different methods}
\label{Fig_time}
\end{center}
\end{figure}

 {This section is dedicated to the evaluation of the running time of all different methods. For this purpose, each method has been executed 100 times with each network and the average running time is reported in Fig.~\ref{Fig_time}. The x-axis lists the networks in ascending order of the number of edges. As seen in the figure, the running time of methods increases as the number of edges increases. LS and ECVC have the first and second highest running time followed by DIL, however, all methods (with the only exception of the rather simple D) appear to perform within one order of magnitude. As expected, the core of ECVC, which is SDC, performs better than ECVC. In fact, SDC's running time does not appear to have a difference in performance compared to the majority of other methods.}

% *********************************************************************************************

\section{Summary}
\label{Section:6}

The problem of identifying users' influence and ranking them in social networks has attracted considerable attention in various research studies; different methods have been proposed for it. The distance to the core of graph is an effective criterion in users' influence. This observation motivated this paper, which has taken into account k-shell and degree of first- and second-order neighbours to propose a composite centrality measure, based on covariance, to determine the importance of each node. A set of experiments were conducted to evaluate the discrimination ability and accuracy of the proposed method. The results obtained showed the superiority of the proposed method in comparison to other similar methods. In future work, somebody may investigate different weights for degree and k-shell and additional experiments. Additional work may also consider the relationship between other structural features and the definition of a centrality measure. Finally, it would be interesting to modify the proposed method so that it can be applied in weighted networks. 

\printbibliography

\begin{table}[ht!]
\centering \footnotesize
\caption{Properties of the networks used in experiments}
\begin{tabular}{ |c|c|c|c|c|c|c| } 
 \hline
 Network & $|V|$ & $|E|$ & MD & $\langle d\rangle$ & $\langle d^{(2)}\rangle$& $CC$\\ 
 \hline
 Karate club (KRT)& 34 & 78 & 17 & 4.5882 & 35.6741 &0.5706\\
 \hline
 Dolphins (DLN) & 62 & 159 & 12 & 5.1290 &34.9032&0.2589 \\
\hline
Jazz musician (JZM) & 198 & 2,742 & 100 & 27.6970 &1070.2400 &0.6174\\
\hline
LFR-200 (L200) & 200 & 1,052 & 16 & 10.5200 & 116.6800&0.4335\\
\hline
Netsciense (NTS) &379  &914 & 34  & 4.8232 & 38.686 & 0.7412 \\
\hline
Facebook Food (FFD) & 620 & 2,092 & 132 & 6.74516& 134.319  &0.3309 \\
\hline
LFR-1000 (L1000)& 1,000 & 10,610 & 98 & 21.2200 &797.5360  &0.4232\\
\hline
Euroroad (EUR) & 1,174 & 1,417 & 10 & 2.4140 & 7.2042&0.0167\\
\hline
Chicago (CHG) & 1,467 & 1,298 & 12 & 1.7696 & 9.5801&0.0000 \\
\hline
Hamsterster (HMS) & 2,426 & 16,631 & 273 & 13.7110& 582.9300 & 0.5380 \\
\hline
Ego-Facebook (FCB) &{4,039} & {88,234} & {1392} & {43.6910} & {4656.3200}&{0.6055}\\
\hline
US Power grid (UPG) & 4,941 & 6,594 & 19 & 2.6691 & 10.3327&0.0801 \\
\hline
LastFM (LFM) &{7,624} & {27,806} & {324} & {7.2943} & {185.4370}&{0.2194}\\
\hline
Pretty Good Privacy (PGP) & 10,680 & 24,316 & 205 & 4.5536 &85.9762&0.2659\\
 \hline
\end{tabular}
\label{Tble_Stats}
\end{table}

\pagestyle{empty}
\begin{landscape}
\begin{table*}[ht!]
\centering 
\caption{DM value for ranking list R by different methods}
\footnotesize
\begin{tabular}{ |c|c|c|c|c|c|c|c|c|c|c|c|c|c|c| } 
\hline
network &  KS& {D} & MDD & Cnc+ & KS-IF & {DIL} & MCDE & LS & HKS & DSR & WKD &{SDC} & {CVC}& ECVC\\ 
 \hline
KRT & 0.1176 &{0.3235}& 0.4412 & 0.7647 & 0.7941 &{0.7059}& 0.7059 & 0.7941 & 0.7941 &0.7647 & 0.7941 &{0.7941}& {0.7941}& \textbf{0.8529}\\
 \hline
DLN & 0.0645 & {0.1935}& 0.3710 & 0.8226 & \textbf{0.9677} & {0.8710}& 0.8710 & 0.9516 & \textbf{0.9677} &\textbf{0.9677} & 0.9032& {\textbf{0.9677}}& {\textbf{0.9677}} & \textbf{0.9677}\\
\hline
JZM & 0.1061 & {0.3131}& 0.6263 & 0.9646 & 0.9646 & {0.9343}& 0.9444& \textbf{0.9747} & 0.9646 & 0.9697 & 0.9697 & {0.9646}& {\textbf{0.9747}}& 0.9697\\
\hline
L200 & 0.0100 & {0.0250}& 0.0850 & 0.3150 & 0.7750 & {0.8000}& 0.5700 & 0.9750 & 0.9500 & \textbf{0.9950} & 0.4600 & {0.5050}& {0.9450}& \textbf{0.9950}\\
\hline
NTS & 0.0211& {0.0554}& 0.1425 & 0.4855 & 0.6807 & {0.3272}& 0.5251 & 0.7071 & 0.6913 & \textbf{0.7150}& 0.6491 & {0.7018}& {0.7071}& 0.7018\\

\hline
 {FFD} & {0.0177}& {0.0661}&  {0.1790} & {0.6968} &  {0.8355} & {0.6710}&  {0.5516} &  {0.8371} &  {0.4648} &  {0.4681} &  {0.7823} & {0.8903}& {\textbf{0.9048}}&  {\textbf{0.9048}}\\
\hline
L1000 & 0.0120 & {0.0430}& 0.1770 & 0.8670 & 0.9870 & {0.9850} & 0.9510 & 0.9960 & 0.9970 & \textbf{0.9990} &0.9490 & {0.9980}& {0.9980}& 0.9980\\
\hline
EUR & 0.0017 & {0.0077}& 0.0187 & 0.0451 & 0.3101 & {0.2078}& 0.1925 & 0.1635 & 0.5954 & 0.7802 & 0.1397 & {0.3296}& {0.6899}& \textbf{0.8722}\\
\hline
CHG & 0.0007& {0.0075}& 0.0109 & 0.0204 & 0.0416 & {0.2154}& 0.0075 & 0.0075 & 0.0723 & 0.0688 & 0.0307 & {0.0300}& {0.0654}& \textbf{0.0995}\\
\hline
HMS & 0.0095& {0.0458}& 0.1467 & 0.6393 & 0.6744 & {0.6195}& 0.6179 & 0.6950 & 0.6810 &0.6962  & 0.6686 & {0.6814}& {0.7042
}& \textbf{0.7045}\\
\hline
 {FCB} &  {0.0238}& {0.0562}& {0.2238} &  {0.9468} &  {0.9549} & {0.9044}&  {0.9369}&  {0.9579} &  {0.9577} & {\textbf{0.9591}} &  {0.9542} & {0.9567}& {0.9582}&  {0.9574}\\
\hline
UPG & 0.0010& {0.0032}& 0.0105 & 0.0306 & 0.2740 & {0.2283}& 0.1813 & 0.3153 & 0.3982 & 0.8089 & 0.01540 & {0.4426}& {0.7766}& \textbf{0.8875}\\
\hline
 {LFM} &  {0.0026}& {0.0129}& {0.0485} &  {0.3831} &  {0.7905} & {0.6422}&  {0.5215}&  {0.8538} &  {0.9136} & {0.9079} &  {0.6321} & {0.9088}& {0.9218}&  {\textbf{0.9226}}\\
\hline
PGP & 0.0024& {0.0078}& 0.0287 & 0.1722 & 0.4677 & {0.3291}& 0.3497 & 0.5724 & 0.6766 & 0.6744& 0.3546 & {0.6495}& {0.7333}& \textbf{0.7450}\\
 \hline
\end{tabular}
\label{Tble_DM}
\end{table*}
\end{landscape}
\pagestyle{plain}

\begin{table}[ht!]
\centering \footnotesize
\caption{The values of the infection threshold $\beta _{th}$ and infection probability $\beta$ considered for different networks}
\begin{tabular}{ |c|c|c||c|c|c| } 
 \hline
 Network & $\beta _{th}$ & $\beta$ &Network & $\beta _{th}$ & $\beta$ \\ 
 \hline
 KRT & 0.129 & 0.15 & EUR & 0.333 & 0.35\\ 
 \hline
 DLN & 0.147 & 0.15 & CHG & 0.185 & 0.2\\ 
 \hline
 JZM & 0.026 & 0.05  &HMS & 0.024 & 0.03\\ 
 \hline
 L200 & 0.18 & 0.2 & FCB &0.009 &0.01\\ 
 \hline
  NTS & 0.125 & 0.15 & UPG & 0.258 & 0.3\\ 
 \hline
  %ELS & 0.006 & 0.01
  FFD &0.0502 & 0.06& LFM&0.039& 0.04\\ 
 \hline
   L1000 & 0.053 & 0.06 & PGP & 0.053 & 0.1\\ 
 \hline
\end{tabular}
\label{Tble_Beta}
\end{table}

\begin{landscape}
\pagestyle{empty}
\begin{table*}[ht!]
\centering\footnotesize
\caption{Value of Kendall tau correlation coefficient between $R$ and $\sigma$ for different networks and methods}
\begin{tabular}{ |c|c|c|c|c|c|c|c|c|c|c|c|c|c|c| } 
 \hline
Network &  KS & {D}& MDD & Cnc+ & KS-IF & {DIL} & MCDE & LS & HKS &DSR & WKD & {SDC}& {CVC}& ECVC\\ 
 \hline
KRT & 0.5544 & {0.6809}& 0.7041 & 0.7647 & 0.7291 & {0.7059} & 0.7112 & 0.7504 & \textbf{0.7754} & 0.7504 & 0.7469 & {0.7718}& {0.7718}& 0.7647\\
 \hline
DLN & 0.5791 & {0.7721}& 0.8038 & 0.8847 & 0.8445 & {0.8096}& 0.8075 & 0.8377 & 0.9112 &0.8995 & 0.8255 & {0.8477}& {0.8837}& \textbf{0.9143}\\
\hline
JZM & 0.7718 & {0.8634}& 0.8907 & 0.9219 & 0.8754 & {0.844}& 0.9291 & 0.8404 & 0.9179 & 0.9444 & 0.9203 & {0.9174}& {\textbf{0.9485}}& 0.9294\\
\hline
L200 & 0.047& {0.5895}& 0.683 & 0.8451 & 0.8320 & {0.6507}& 0.6975 & 0.5800 & 0.8107 & 0.7386& 0.7762& {0.7832}& {0.8712}& \textbf{0.8847}\\
\hline
NTS & 0.5018 & {0.5681}& 0.5886 & 0.8263 & 0.8171 & {0.5692}& 0.6159 & 0.7870 & 0.8823 & 0.8243 & 0.6852 & {0.7794}& {0.8830}& \textbf{0.8952}\\
\hline
 {FFD}& {0.5657}& {0.5490}& {0.5699}& {0.7907}& {0.7704}& {0.5733}& {0.5817}& {0.7585}& {\textbf{0.8272}}& {0.8077}& {0.6903}& {0.7309}& {0.8036}& {0.8226}\\
\hline
L1000 & 0.5929 & {0.5076}& 0.5479 & 0.7792 & 0.7647 & {0.5092}& 0.6078 & 0.7292 & \textbf{0.7850} & 0.7584 & 0.7134 & {0.7148}& {0.7794}& \textbf{0.7850}\\
\hline
EUR & 0.3993& {0.4811}& 0.5884 & 0.7673 & 0.7697 & {0.6436} & 0.6540 & 0.5976 & 0.8142 & 0.7611& 0.7015& {0.6945}& {0.7900}& \textbf{0.8318}\\
\hline
CHG & 0.0000 & {0.1657}& 0.1669 & 0.5433 & 0.5669 & {0.4194}& 0.1681 & 0.1657 & 0.5941 & 0.4618& 0.5414& {0.5065}& {0.5752}& \textbf{0.6125}\\
\hline
HMS & 0.6836& {0.6959}& 0.7007 & 0.8266 & 0.8245 & {0.7142}& 0.7085 & 0.8214 & 0.8347 & 0.8047& 0.7730& {0.7905}& {0.8360}& \textbf{0.8366}\\
\hline
 {FCB}& {0.6970}& {0.6659}& {0.6844}& {0.8599}& {0.8501}& {0.6657}& {0.6933}& {0.7634}& {0.8554}& {0.7743}& {0.7569}& {0.7646}& {0.8288}& {\textbf{0.8624}}\\
\hline
UPG & 0.3359& {0.4715}& 0.5071 & 0.7147 & 0.7246 & {0.5134}& 0.5447 & 0.5047 & \textbf{0.7974} & 0.7262& 0.6196& {0.6377}& {0.7507} & 0.7972\\
\hline
 {LFM}& {0.6161}& {0.5897}& {0.6036}& {0.8370}& {0.8104}& {0.5948}& {0.6270}& {0.7762}& {0.8537}& {0.8150}& {0.7455
}& {0.7409}& {0.8245}& {\textbf{0.8560}}\\
\hline
PGP & 0.4073& {0.4252}& 0.4361 & 0.7144 & 0.6821 & {0.4286}& 0.4490 & 0.6798 & 0.7344 &0.7199& 0.6214& {0.5736}& {0.6856}& \textbf{0.7348}\\
 \hline
\end{tabular}
\label{Tble_Tau}
\end{table*}
\end{landscape}
\pagestyle{plain}

\end{document}